\documentclass{article}

\usepackage{arxiv}

\usepackage[utf8]{inputenc} 
\usepackage[T1]{fontenc}    
\usepackage{hyperref}       
\usepackage{url}            
\usepackage{booktabs}       
\usepackage{amsfonts}       
\usepackage{nicefrac}       
\usepackage{microtype}      
\usepackage{lipsum}
\usepackage{graphicx}
\graphicspath{ {./images/} }
\usepackage{subcaption}

\usepackage{amsmath}

\usepackage{hyperref}
\hypersetup{
    colorlinks=true,
    linkcolor=blue,
    filecolor=magenta,      
    urlcolor=blue,
}

\title{Traffic Performance Score for Measuring the Impact of COVID-19 on Urban Mobility}

\author{ Zhiyong Cui, Meixin Zhu, Shuo Wang, Pengfei Wang, Yang Zhou, Qianxia Cao, Cole Kopca, Yinhai Wang \thanks{Corresponding Email: yinhai@uw.edu}  \\
  Department of Civil and Environmental Engineering\\
  University of Washington\\
  Seattle, WA 98195 \\
}

\begin{document}
\maketitle

\begin{abstract}
Measuring traffic performance is critical for public agencies who manage traffic and individuals who plan trips, especially when special events happen. The COVID-19 pandemic has significantly influenced almost every aspect of daily life, including urban traffic patterns. Thus, it is important to measure the impact of COVID-19 on transportation to further guide agencies and residents to properly respond to changes in traffic patterns. However, most existing traffic performance metrics incorporate only a single traffic parameter and measure only the performance of individual corridors. To overcome these challenges, in this study, a Traffic Performance Score (TPS) is proposed that incorporates multiple parameters for measuring network-wide traffic performance. An interactive web-based TPS platform that provides real-time and historical spatial-temporal traffic performance analysis is developed by the STAR Lab at the University of Washington. Based on data from this platform, this study analyzes the impact of COVID-19 on different road segments and the traffic network as a whole. Considering this pandemic has greatly reshaped social and economic operations, this study also evaluates how  COVID-19 is changing the urban mobility from both travel demand and driving behavior perspectives.
\end{abstract}


\section{Introduction}

Traffic performance measurement is incredibly beneficial to both transportation agencies, who use this data as a source of strategy for operations and management, and to the general public, as traffic patterns impact daily life. Many classical traffic performance metrics, including speed, volume, travel time, etc., provide relatively intuitive information about the traffic status of a roadway segment or a specific corridor. However, there are few existing metrics that can indicate network-wide traffic performance. This is especially true when the traffic network scale is large enough to cover an entire city. Further, because most existing traffic metrics only measure one element of traffic stream characteristics, they have difficulty distinguishing complicated traffic scenarios. For example, two roadways with the same traffic speed, one with a high volume and one low volume, at two different times, indicates two entirely different traffic scenarios. Therefore, to overcome the shortcomings of these classical traffic metrics, this study attempts to develop a measurement metric to indicate the network-wide traffic performance in a more comprehensive way.

The first COVID-19 case in the United States was reported in January 2020. Since that time, cases have occurred in all 50 U.S. states. By the end of May, 2020, the U.S. has more than 1,800,000 confirmed cases and more than 100,000 deaths \cite{DavidWelna2020}. This pandemic has not only dramatically influenced the social and economic activities, but also hugely affected the transportation system. Traffic patterns in most cities have been entirely reshaped since January, due to COVID-19. For example, historical traffic data indicates that the traffic volume in the Greater Seattle, Washington, area started to decline four weeks before the Washington State's stay-at-home order on March 23. On the week the stay-at-home order was released, traffic volume dropped significantly. Many other aspects of the traffic have also been affected by COVID-19, such as dramatic decreased in public transportation, traffic safety and traffic congestion. Since the pandemic reached the U.S., its influence on transportation has been extensively studied by agencies, researchers, and private companies. Although aspects of the impact have been quantitatively and qualitatively analyzed based on massive data, few studies have been able to measure the impact of COVID-19 on urban mobility from both a road segment level and a traffic network level. One of the targets of this study was to build and release a big data-based traffic performance measurement platform to assist in COVID-19 related analysis.

\begin{figure*}[t]
  \centering
  \fbox{\includegraphics[width=0.95\textwidth]{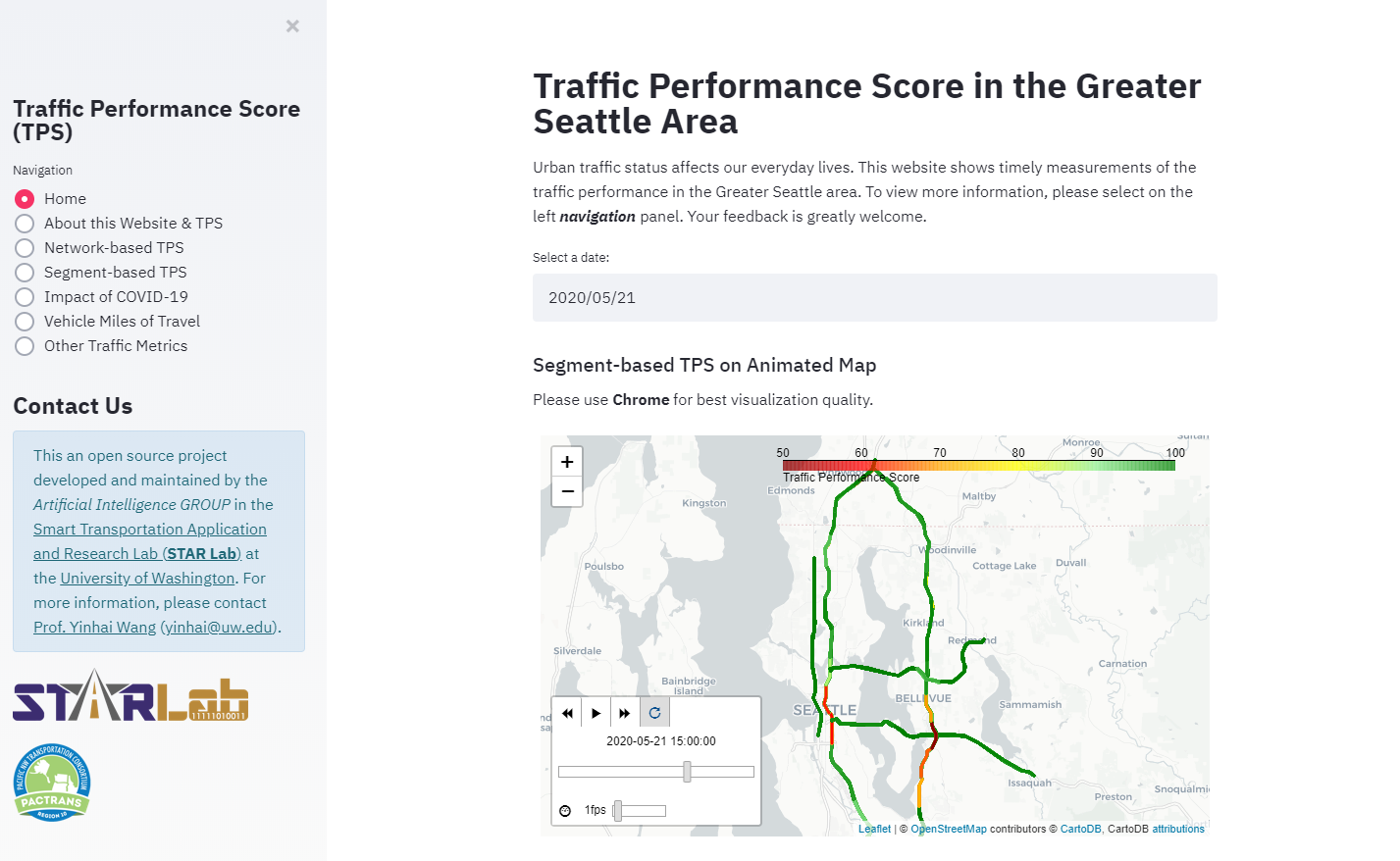}}
  \caption{The web interface of the traffic performance score platform.}
  \label{fig:TPS}
\end{figure*}

To observe changes in traffic patterns due to COVID-19, interested parties can utilize public map services, such as Google Map and Apple Map, or review performance metrics from public agencies. However, in most cases, these traffic performance measurements only provide fine-grained (i.e. road segment-level) real-time traffic information or coarse-grained (i.e. network-level) long-term traffic analysis. During the COVID-19 pandemic, segment-level, long-term analysis and network-wide, real-time traffic performance information are vital to guide personal travel activities and support agency transportation operation management. To that end, this study analyzes both segment-level and network-level traffic performance based on a proposed traffic performance score and other existing traffic metrics, such as vehicle miles of travel (VMT). To measure the impact of COVID-19 on traffic, both short-term and long-term mobility patterns are analyzed.

The contribution of this paper can be summarized as follows:
\begin{enumerate}
    \item A traffic performance score (TPS) for measuring network-wide traffic is proposed. The TPS is also implemented on a public accessible TPS platform \footnote{TPS Platform: \url{ http://tps.uwstarlab.org/}} for the Greater Seattle area, as shown in Figure \ref{fig:TPS}.
    \item The network-wide traffic performance under the influence of COVID-19 in the Greater Seattle area is analyzed based on the proposed TPS and other metrics. The variations of TPS and VMT show that responses to the pandemic greatly affect urban traffic.
    \item The traffic changes of different road segments in the traffic network are analyzed from both travel demand and driving behaviour perspectives, to investigate how COVID-19 is reshaping the urban mobility. 
\end{enumerate}

In the following sections, this paper first introduces the existing traffic performance metrics and COVID-19 related analysis in Section 2. Section 3 introduces the proposed TPS and describes the design principles and techniques used to develop the TPS platform. Then, the traffic performance changes in response to COVID-19 from network-scale perspective is presented in Section 3. Finally, road segment level traffic changes are analyzed in Section 4 to shed light on the changes to urban mobility due to COVID-19.

\section{Literature Review}

\subsection{Existing Traffic Performance Metrics}
Traffic performance metrics are based on various essential traffic parameters such as speed, travel time, delay, volume, density, and vehicle miles traveled (VMT). This section summarizes several traditional traffic performance metrics and discusses some recent related studies.

\begin{itemize}
    \item \textbf{Travel Time Index (TTI) \cite{bharti2013performance}}: compares peak (investigated) travel time and free-flow travel time, as defined by:\\
    \begin{equation}
    TTI = \frac{\text{Peak Period Travel Time}}{\text{Free-Flow Travel Time}}
    \end{equation}
    A large TTI value indicates a state of congested traffic. One of the limitations of TTI is that, while it has a lower bound, it does not have any upper bound. 
    
    \item \textbf{Level of Service (LOS)}: one of the most popular traffic performance metrics. The Highway Capacity Manual (HCM) \cite{manual2016guide} divides traffic operations into six levels based on traffic density: A (free flow), B (reasonably free flow), C (stable flow), D (approaching unstable flow), E (unstable flow), and F (forced or breakdown flow). While LOS is easy to understand, it also has the following limitations: (1) it is hard to measure traffic density; and (2) LOS is not a continuous metric.
    
    \item \textbf{Lane Mile Duration Index (LMDI) \cite{rao2012measuring}}: calculated by summing over the product of congested segment length and congestion time, as shown in the following equation: 
    \begin{equation}
        LMDI = \sum_{i = 0}^m L^i*t^i
    \end{equation}
    where $L^i$ is the length of the segment $i$, $t^i$ is the congestion time of segment $i$, and $m$ is the total number of segments. One of the limitations of this metric is that it is not scaled and thus can not be used to compare the performances of different traffic networks, due to the fact that the total number of segments may be different in each network. 
\end{itemize}

Recently, He et al. \cite{he2016traffic} proposed a framework to evaluate traffic performance of road segments and network. This basic metric is named speed performance index which simply averages speed divided by the free flow speed. To aggregate segment-based metrics to network scale, segment lengths were used as weights for aggregation. Lee and Hong \cite{lee2014congestion} proposed a traffic performance metric called Traffic Congestion Score (TCS) based on link travel speeds. The TCS is calculated as:
    \begin{equation}
        TCS_i = \begin{cases} 
      {100} \% & v_i \leq 0 \\
      0\% & v_i \geq v_l\\
      (1-\frac{v_i}{v_l})\times 100\% & otherwise
   \end{cases}
    \end{equation}
    where $v_i$ is the average travel speed of link $i$, $v_l$ is the speed limit. 
    
To summarize, there are several important limitations for the existing traffic performance metrics: (1) they consider only single, simple metrics such as speed or travel time, which cannot fully capture the state of traffic; (2) most of the performance metrics can only investigate individual segments rather than corridors or entire traffic networks; and (3) the above summarized methods have not incorporated VMT to reflect the total travel demands. In this work, to overcome these limitations, a metric to measure network-wide traffic performance is proposed.

\subsection{Impact of COVID-19 on Transportation}

Since COVID-19 first appeared, most cities around the world have implemented varying degrees of lock-down policies for their residents, which has had a direct impact on human mobility. Accordingly, many recent studies have sought to investigate the impact of COVID-19 on transportation. Further, many organizations and companies, such as INRIX, TomTom, Google, and Mapbox, have provided significant amounts of data and tools for COVID-19 related research. 


Most research that investigates COVID-19's impacts on human mobility is based on traffic volume or VMT. Daniel\cite{Daniel2020} shows how traffic data can help inform policy decisions by visualizing residual mobility along side new cases of COVID-19, based on data from TomTom. This work asserts that traffic data suffices as a measure of the policy because of its time and space attributes. Another report \cite{Andy2020} investigating the effect of the COVID-19 lockdown on mobility reveals that the top ten cities with the highest traffic reductions worldwide all saw reductions of over 80\%. Shi et.al. \cite{shi2020temporal} examined the time-lagged effect between outbound traffic from Wuhan, China, and the status of the COVID-19 pandemic.

Other COVID-19-related research  has focused on traffic safety metrics such as incident numbers and implications of social distancing. C2SMART, \cite{Gao2020} University Transportation Center at New York University, released a white paper on the impact of COVID-19 on the transportation system, which uses the New York metropolitan area as a case study to analyze transit ridership, subway turnstile entries, traffic on bridges and tunnels, travel time, and the number of crashes during COVID-19. Zhang et al. \cite{zhang2020interactive, UMPlatform} developed an interactive COVID-19 Impact Analysis Platform introducing nine metrics related to human mobility and social distancing, and fifteen metrics on COVID-19 and health at the nation-, state-, and county-levels. The social distancing index introduced in this research provides meaningful support for decisions on travel and social distancing policies.

Other recent research explores the impact of COVID-19 on traffic congestion. A congestion index \cite{Sean2020} based on the traffic index from TomTom was introduced to compare congestion during peak hours in 2019 and 2020. After the COVID-19 lockdown policies were announced, the congestion indices of cities around the world declined to varying degrees. David \cite{David2020} also proposed a traffic congestion index defined as the percent difference between drive times and the average baseline drive time. Based on the data collected from Google Maps Distance Matrix API, the daily and hourly congestion levels were analyzed. This study found that reduced congestion is also related to the weather. Finally, Maricopa Association of Governments in Arizona, released a report \cite{MAG2020} tracking traffic congestion during COVID-19 stay-at-home restrictions based on traffic volumes from INRIX data.

Based the work described above, it appears traffic metrics are affected by the COVID-19 pandemic from multiple perspectives. Based on this assertion other research has investigated changes in travel patterns due to changing traffic scenarios. Rachel et.al \cite{Rachel2020} analyzed changes in public transit, bicycling and scootering, air travel, driving, ride-hailing, and delivery vehicles based on stay-at-home policies. Michelle \cite{Michelle2020} compared traffic patterns from January 13 to April 27 in Seattle based on GPS mobile device data collected by Mapbox. This work finds that traffic on major highways and freeways declined, while traffic on neighborhood streets increased. However, the methods employed were not sufficient to show detailed data and the trend of the change.

Beyond the present, COVID-19 is likely to have long-term effects even as cases and deaths wane. INRIX published a United States National Activity Re-Emergence Map to track the trend of city activities recovering from COVID-19 \cite{INRIX2020}. Hu et.al analyzed the travel mode shift effect on traffic caused by COVID-19 \cite{hu2020impacts, Huplatform}. This work predicted travel time, post pandemic, across the United States with an interactive Rebound Calculator. It analyzed shifts from transit commuters to single occupancy, from carpool to single occupancy, job loss, and shifts to remote work. The findings indicate that the long term effects of the pandemic on travel may result in increased travel time as communities reopen. Alessio et.al.\cite{tardivo2020covid} published work on COVID-19's impact on railway system. This paper illustrates the effect on global travel manners and the possible rebound of traffic activities after the COVID-19.

In summary, many studies have analyzed the impact of COVID-19 on transportation from various aspects, including impacts on average speed, traffic volume, traffic congestion, traffic patterns, the future effects and so on. However, few researchers have developed indices to intuitively show traffic variation. Those who have developed indices, narrowly focused on one aspect of the impacts but not comprehensive changes to the network. Thus, this study proposes a traffic performance score to measure traffic conditions, and also analyzes how COVID-19 is reshaping urban mobility from road segment and network perspectives.

\section{Traffic Performance Score}

Performance monitoring is critical for roadway operations, including real-time applications and operational planning, and transportation planning. According to \cite{turner2004lessons}, travel time is the basis for defining mobility-based performance measures. To that end, many performance measures have been designed, such as average travel speed, travel time, travel rate index, and delay per VMT. However, most of the existing performance measures are road segment or trip-based measures. These existing metrics cannot measure performance over a complicated road network. In this section, this paper outlines a traffic performance score to measure performance from the road network perspective.

\subsection{Data Source}

\begin{figure}
  \centering
  \fbox{\includegraphics[width=0.7\textwidth]{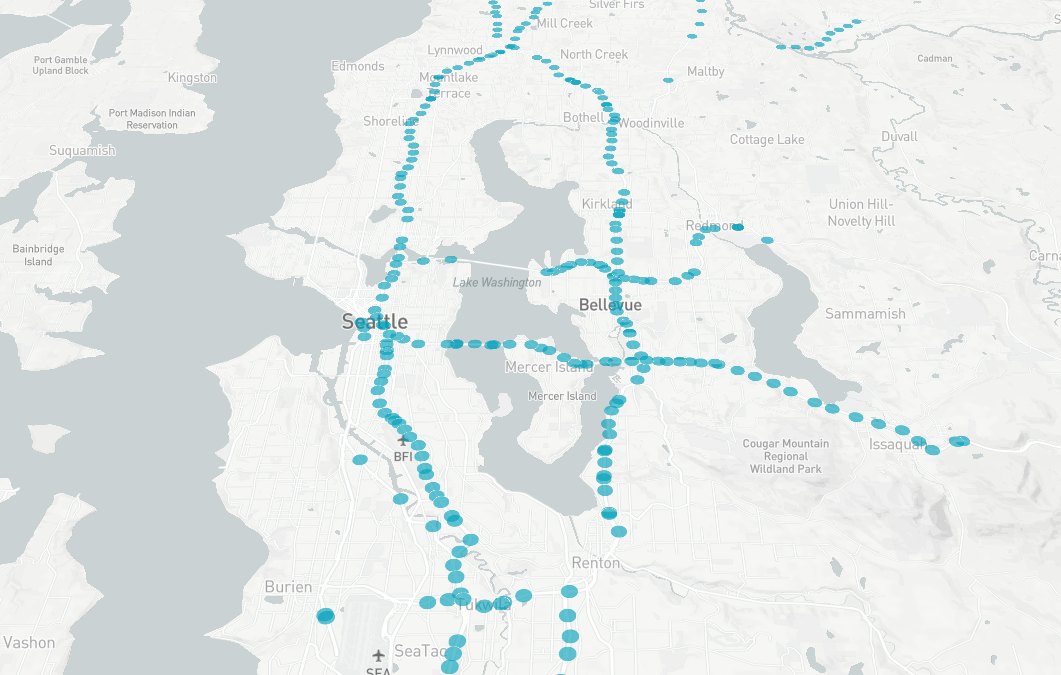}}
  \caption{The spatial distributions of the loop detector sensors in the Greater Seattle area.}
  \label{fig:data}
\end{figure}

The TPS is calculated based on data collected from roughly 8000 inductive loop detectors deployed on the freeway network in the northwest region in Washington State. The freeway network mainly includes several major freeways, such as I-5, I-90, I-99, I-167, I-405, and SR-520. The raw data comes from the online Washington State Department of Transportation (WSDOT) Traffic Data Archive Folder \footnote{Website: \url{ http://data.wsdot.wa.gov/traffic/}}. Representative detectors are shown by blue dots in the map in Figure \ref{fig:data}.

The raw data contains lane-wise speed, volume, and occupancy (density) information collected by each loop detector. Each detector's meta data includes detector category, route, milepost, director, direction, address. Based on the consecutive detectors' location information, freeways can be separated into tiny road segments, each of which contains one loop detector per lane. A road segment's length is then considered as the corresponding detector's covered length. The time interval of the data is one-minute.

Other data sources includes public agencies, public accessible datasets, and official COVID-19 related datasets. In this study, the COVID-19 datasets come from the Wikipedia page of COVID-19 pandemic in Washington \footnote{\url{https://en.wikipedia.org/wiki/COVID-19_pandemic_in_Washington_(state)}}.

\subsection{Definition}

Normally the physical properties of road segments are invariant, but the traffic stream parameters on each segment, including volume (Q), speed (V), and density (K), keep changing. In order to measure the traffic performance of the traffic network, these traffic parameters at each road segment should be taken into account. Since volume, speed, and density are all related to each other, only two of them need to be incorporated. Thus, volume and speed, which are more general and intuitive parameters are adopted in the design of TPS. The length ($L$) of the road segment is also taken into consideration by multiplying $L$ with the volume $Q$, which ($Q \cdot L$) basically represents the vehicle miles of travel (VMT) of the road segment. Then, the TPS is defined as:

\begin{equation}
TPS_{t} = \frac{\sum_{i=1}^n V_t^i \cdot Q_t^i \cdot L^i}{\sum_{i=1}^n V_f \cdot Q_t^i \cdot L^i} \times 100\%
\end{equation}

\noindent where $V_t^i$ and $Q_t^i$ represent the speed and volume of each road segment $i$ at time $t$. $L^i$ is the length of $i$-th detector's covered road segment. $V_f$ is the default free-flow speed. 

In this way, the TPS is a value ranging from 0\% to 100\%. Overall network-wide traffic condition is best when the TPS is 100\% and worst when TPS is 0\%.

\begin{figure*}
  \centering
  \includegraphics[width=0.95\textwidth]{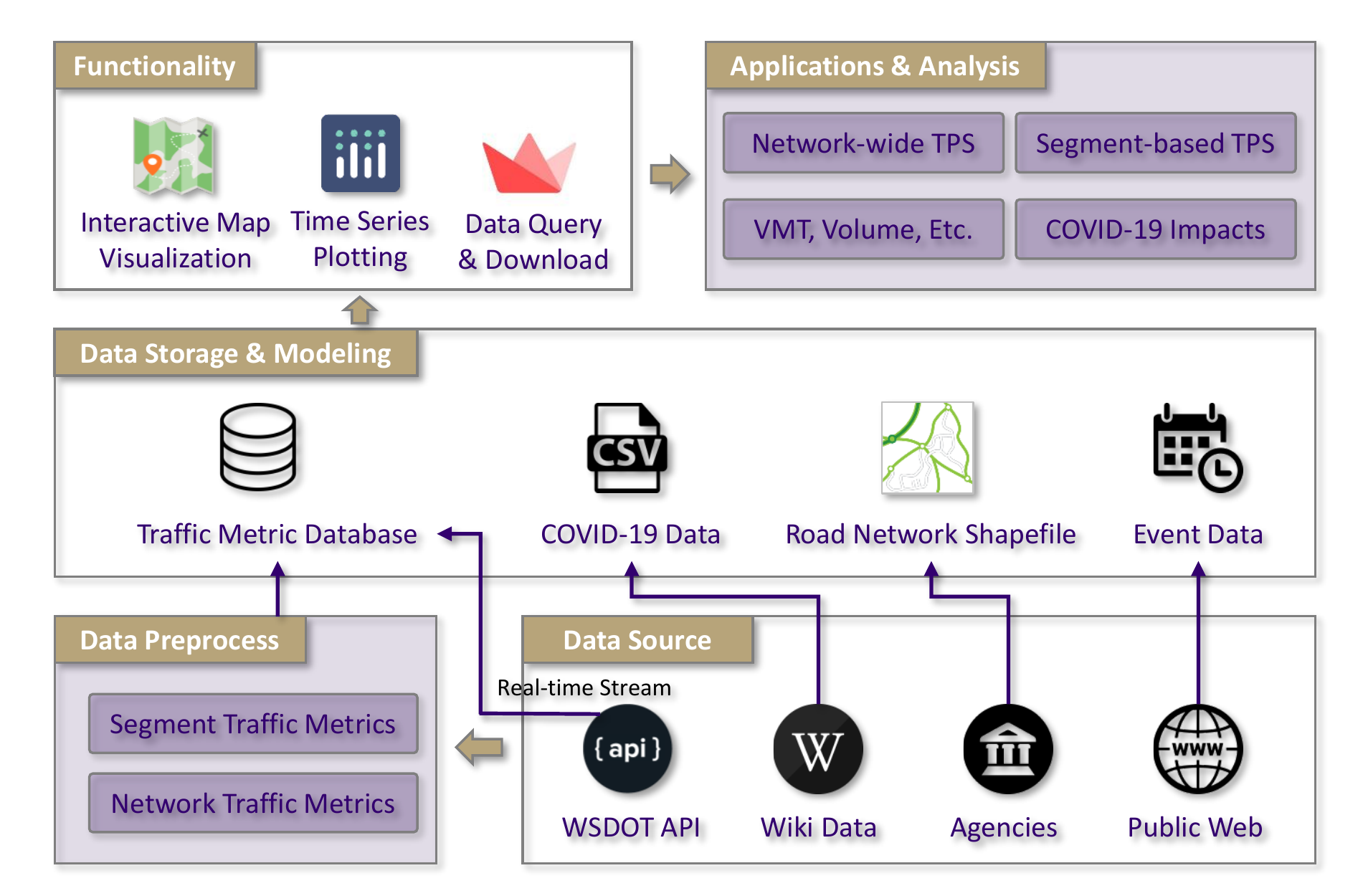}
  \caption{Architecture of Traffic Performance Score Platform}
  \label{fig:architecture}
\end{figure*}

\subsection{System Design}

Based on multi-source real-time data, the network-wide and segment-level traffic performance measurement analytical functions and applications are implemented on an interactive publicly accessible web-based traffic performance score platform (\url{ http://tps.uwstarlab.org/}). This platform integrates real-time data streaming, data source pre-processing, data storage, data modeling, and analytical data visualization using the Streamlit tool \footnote{\url{https://www.streamlit.io/}} which is typically employed to build highly interactive machine learning and data science web applications. The architecture of the platform contains three primary layers is shown in Figure \ref{fig:architecture}.

The first layer is the data streaming and pre-processing layer. The real-time inductive loop detector data from WSDOT API is cleaned, archived, and processed in parallel to calculate network-wide and segment-level traffic performance metrics. The second layer of the platform is mainly responsible for storing traffic raw data and calculated intermediate traffic performance metrics into the database, and building the connections between different datasets based on their spatial-temporal properties. Specifically, each road segment's traffic data is linked to the segment's geospatial properties based on the roadway segment's GIS information, i.e. the WSDOT 24K map \footnote{\url{https://www.wsdot.wa.gov/mapsdata/geodatacatalog/maps/NOSCALE/DOT_TDO/LRS/WSDOT_LRS.htm}}. In the third layer, advanced interactive spatial-temporal data visualization tools are employed to demonstrate a variety of network- and segment-level traffic performance measurement analysis results.

\section{Traffic Changes in Response to COVID-19}

In this section, the impact of COVID-19 on traffic changes reflected by different traffic performance metrics is discussed. Firstly, government responses to COVID-19 in Washington State are investigated, as well as some responses from large employers in and around Seattle, listed as follows:
\begin{itemize}
    \item March 6: Major tech companies ask Seattle employees to work from home. Amazon and Facebook shut down individual offices as well.
    \item March 9: University of Washington (UW) suspends on-site classes and finals.
    \item March 13: Washington state governor announces statewide school closures, expansion of limits on large gatherings.
    \item March 16: Washington state governor announces statewide shutdown of restaurants, bars and expanded social gathering limits.
    \item March 23: Washington state governor announces "Stay Home, Stay Healthy" order.
    \item April 2: Washington state governor extends "Stay Home, Stay Healthy" through May 4.
\end{itemize}

\begin{figure}
  \centering
  \includegraphics[width=0.7\textwidth]{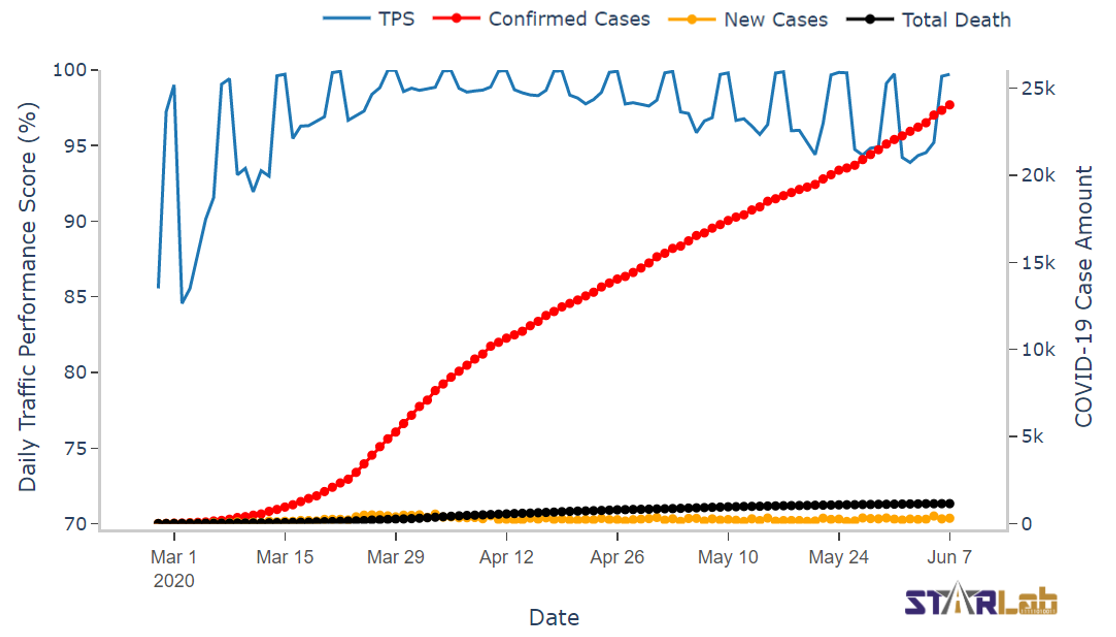}
  \caption{COVID-19 cases in Washington State and network-wide traffic performance score in the Greater Seattle area.}
  \label{fig:covid19}
\end{figure}

\subsection{Impact of COVID-19 on TPS}

\begin{figure*}[]
  \centering
  \includegraphics[width=0.95\textwidth]{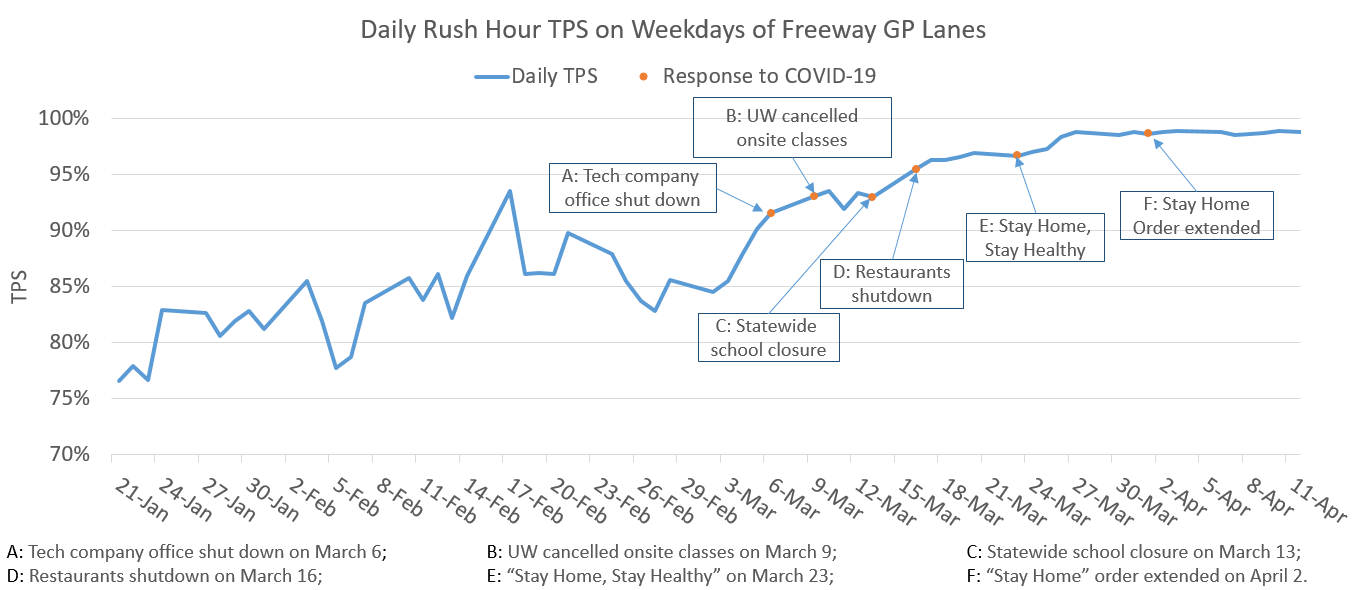}
  \caption{Weekday rush hour TPS of freeway general purpose (GP) lanes.}
  \label{fig:covid19tps}
\end{figure*}

Since the first death from COVID-19 in the U.S. was announced in Kirkland, an east-side suburb of Seattle, on February 29, 2020, the traffic pattern in the Greater Seattle area has gradually changed along with the public response to COVID-19. COVID-19 cases and the calculated daily network-wide TPS are displayed on the TPS platform, as shown in Figure \ref{fig:covid19}. The red, yellow, and black dotted lines shows cumulative confirmed cases, daily new cases, and cumulative deaths, respectively. Although the values of TPS on weekends generally approach  100\%, it can be observed that the TPS dramatically increased on weekdays in March. The weekday TPS reached its peak in early April and then gradually decreased, though it still had not recovered to its pre-COVID-19 status as of early June 2020. 

In Figure \ref{fig:covid19tps}, the government and major employer responses to the pandemic are illustrated in order to demonstrate their influence on the TPS. Figure \ref{fig:covid19tps}, only shows the weekday rush hour TPS on freeway general purpose (GP) lanes. Based on the preliminary analysis, the average weekday TPS during the rush hour was 83.5\% before COVID-19 (from Jan. 21 to Feb. 28). By March 2, the TPS began to increase, implying that residents started to decrease or stagger their traveling activities. After tech companies shut down offices on March 6 and UW moved classes online on March 9, the TPS increased to 92.7\%. The orders of statewide school closure and restaurants/bars shutdown led the further increase of the TPS to 95.9\%. Finally, once the governor's stay-home order was announced, the TPS leveled out at its highest point around 98\%.

\subsection{Impact of COVID-19 on VMT}

\begin{figure*}[]
  \centering
  \includegraphics[width=0.95\textwidth]{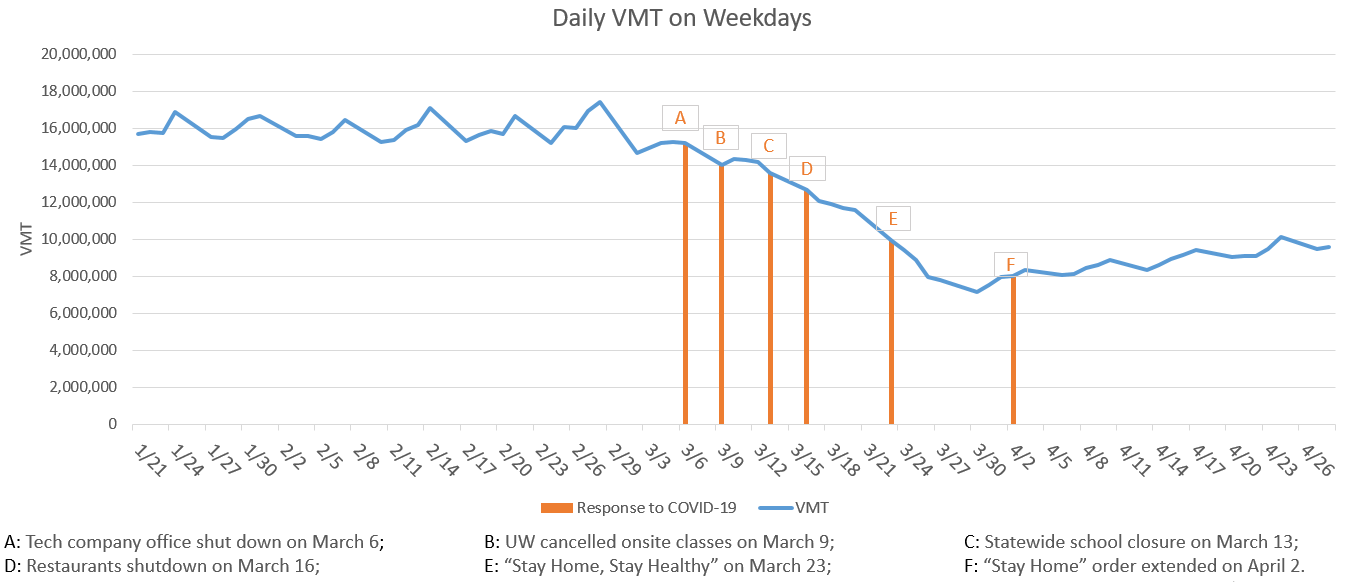}
  \caption{Daily VMT changes on weekdays.}
  \label{fig:covid19vmt}
\end{figure*}

\begin{figure*}
  \centering
  \includegraphics[width=0.95\textwidth]{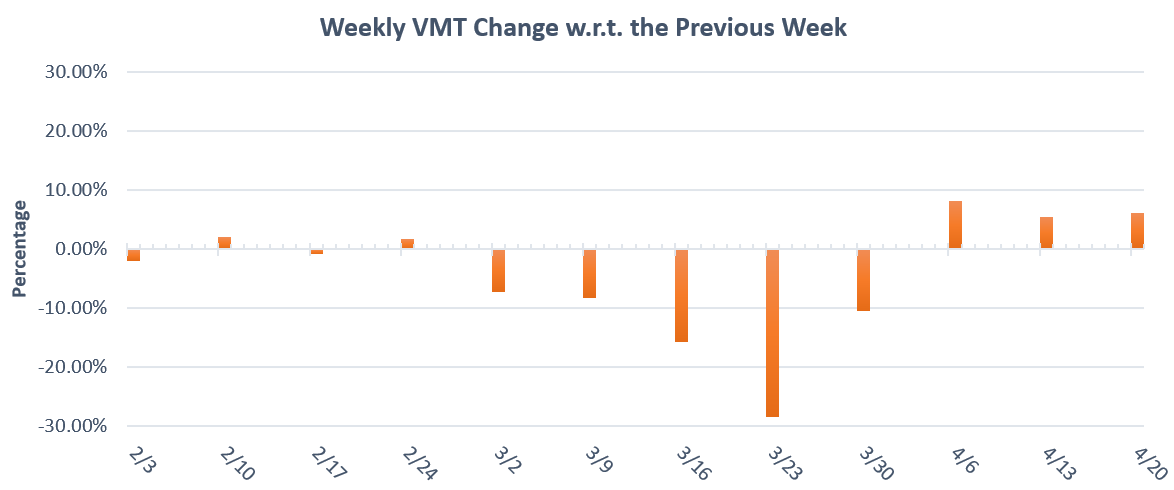}
  \caption{Weekly VMT changes with respect to that of the previous week.}
  \label{fig:vmtweekly}
\end{figure*}

VMT is another important metric measuring the amount of travel for all vehicles in a geographic region over a given period of time. Given the lane-level loop detector data, the VMT of the freeway system in the Greater Seattle area can be calculated. Figure \ref{fig:covid19vmt} shows the variations of daily VMT on weekdays before and during the presence of the COVID-19 pandemic. Before the outbreak, VMT on the freeway system in the Greater Seattle area was steadily around 16 million per weekday. In correlation with rises in new cases and deaths, and the government and large employer responses, the VMT began to decrease. When Gov. Inslee announced the stay-home order on March 23, the daily VMT dropped by more than half compared to the daily VMT before COVID-19. On March 30, the daily VMT reached the lowest value around 7.1 million.

While the variations of daily VMT are small, weekly VMT clearly reflects the impact of the government and large employer responses to COVID-19. Figure \ref{fig:vmtweekly} shows the percentage of the weekly VMT change with respect to the week prior from early February to late April 2020. It is apparent that weekly VMT remained relatively constant before March, but began to noticeable decline in early March. When the stay-home order was announced in the week starting from March 23, the weekly VMT decreased by its largest single drop, showing a 28.5\% decline w.r.t. to that of the previous week. It continued to decline until early April. The weekly VMT started to increase from the week of April 6, but the growth rate is relatively low compared to the rate of decrease through the month of March, staying around 7\%.

\begin{figure}
  \centering
  \includegraphics[width=0.8\textwidth]{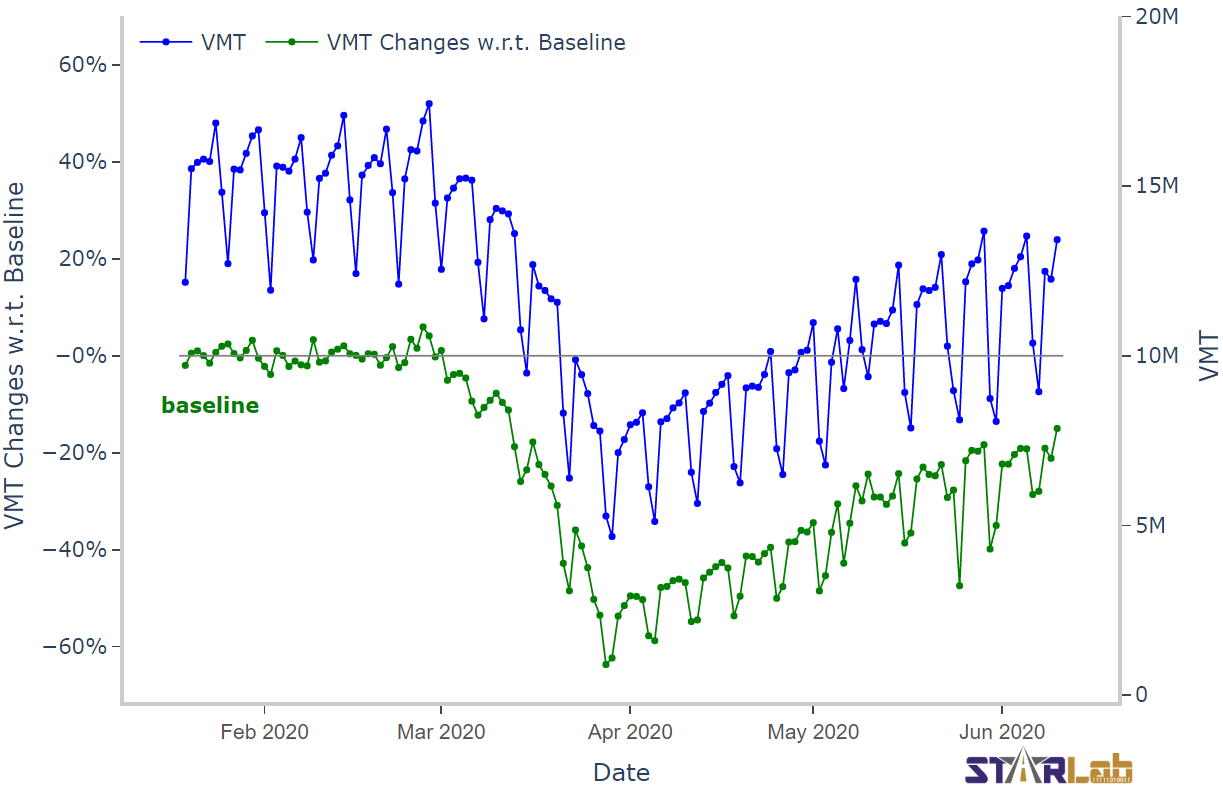}
  \caption{VMT changes with respect to baseline. The baseline is the averaged VMT on each day of the week from Jan. 19 to Feb. 22, 2020.}
  \label{fig:vmtbaseline}
\end{figure}

The TPS platform also provides a comparison function to contrast the daily VMT with a baseline, i.e. the averaged VMT on each day of the week during a specific period. Figure \ref{fig:vmtbaseline} shows the daily VMT by the blue line and VMT changes w.r.t. the baseline from January 19 to February 22 by the green line. The VMT variation in Figure \ref{fig:vmtbaseline} shows a similar trend to that in Figure \ref{fig:covid19vmt}, where the VMT dropped rapidly at the beginning of COVID-19 and gradually increased after the stay-home order was announced. However, the green line in Figure \ref{fig:vmtbaseline} also shows greater detail with regard to weekdays vs. weekends. Namely, weekends showed a much larger drop in the VMT as compared to weekdays, which suggests that weekday traffic during COVID-19 was closer to normal, pre-COVID-19 traffic as compared to weekends. One potential explanation may be that non-essential trips over the weekend  reduced significantly during COVID-19, but the necessary trips taken on weekdays for supporting essential social and economic operations, remained more consistent with pre-COVID-19 patterns.


\begin{figure*}[]
  \centering
  \includegraphics[width=\textwidth]{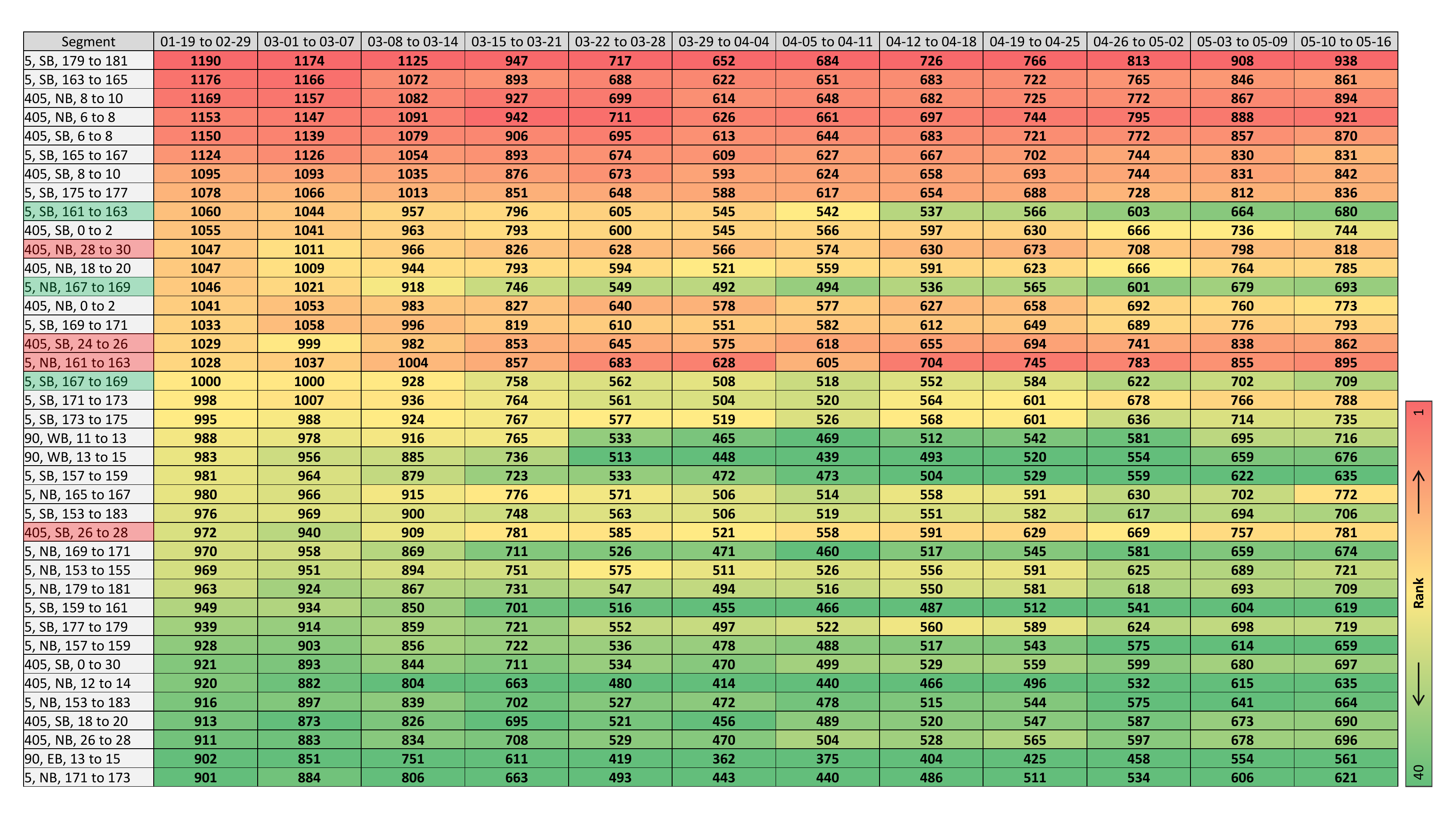}
  \caption{Road segment rank based on volume per lane per hour (VpLpH) of the freeway network in the Greater Seattle area. The header shows the time periods, in which the first period (01-19 to 02-29) is considered as the baseline. The values in table cells indicate a segments' VpLpH during each respective period.}
  \label{fig:segmentrank}
\end{figure*}

\section{How is COVID-19 Reshaping Urban Mobility?}

Although TPS and VMT reflect different aspects of traffic performance characteristics, they both measures the network-wide traffic performance. The traffic performance changes reflected by these two metrics show similar trends during the COVID-19 pandemic. However, because road segments in the traffic system distribute vehicles in different spatial regions and have differing functional properties, the impacts on road segments might differ from one another. Thus, to better get at more detailed traffic changes, it is also beneficial to analyze road segment-level traffic performance.

In the Greater Seattle area, four major connected freeways, including I-5, I-90, I-405, and SR-520, form a large network connecting the City of Seattle, the City of Bellevue, the Seattle-Tacoma International Airport, and many other important functional areas. To analyze segment-level traffic performance, this traffic network, consisting of more than 180 miles roadways, is separated into small road segments with the length of two miles. In this way, traffic performance metrics for each segment can be calculated. In this section, the traffic changes of different road segments are analyzed to investigate how COVID-19 is reshaping urban mobility.

\subsection{Reshaping Travel Demand}

As articulated in the previous section, the traffic volume over the entire freeway network decreased. However, because each road segment serves different functions, the volume of some segments during COVID-19 reduced more dramatically, while others did not. To quantify the relative volume changes of road segments during COVID-19, the volume per lane per hour (VpLpH) of road segments is caluclated during a specific period and then ranked. Figure \ref{fig:segmentrank} shows 40 segments, using different colors to visual their respective rank during the time period from January to May 2020. The value in each cell represents the VpLpH of the corresponding segment during a specific period, and the headers show the time periods, where the first period, or column, (from Jan. 19 to Feb. 29) is considered the baseline. It can be observed that the ranking is in order from red to green in the first column (the baseline) but then does not stay constant throughout each time period. This shows that although the VpLpH of all the segments decreased, the ranks did not remain constant. The ranks of several segments noticeably increased while the ranks of several others decreased, which is why columns other than the baseline column do not maintain the clear gradient from red to green. The relative increase in VpLpH suggests an increased importance of those road segments marked with red color. Conversely, those road segments (rows) that move to a darker green suggest those segments are less important during the COVID-19 pandemic. The change of ranks, or say the relative importance of road segments, results due to the change of travel demand under the influence of COVID-19. The stay-home order caused many residents to reduce daily travel. But the trips made by the people who maintain essential services likely did not see the same drop. As Figure \ref{fig:segmentrank} shows, the decreased travel demand is not evenly distributed across the traffic network. This presents an opportunity to further analyze the functional role of each road segment. Interestingly, the VpLpH changes of road segments at I-5 from milepost 161 to 163 in each direction, which located between the City of Seattle and the Sea-Tac Airport, flip; where the southbound segment gets greener over time, suggesting it is of less importance during COVID-19, while the northbound segment gets redder, suggesting it is of more importance. The rank of the northbound segment increased from around 20 to top 2, but the rank of the southbound segment gradually decreased from around 10 to around 30. These significant rank changes reflect the shift in the respective roles of each road segments on the network during COVID-19 as compared to normal traffic scenarios.

\subsection{Reshaping Driving Behaviour}

\begin{figure*}[]
     \centering
     \begin{subfigure}[b]{0.48\textwidth}
         \centering
         \includegraphics[width=\textwidth]{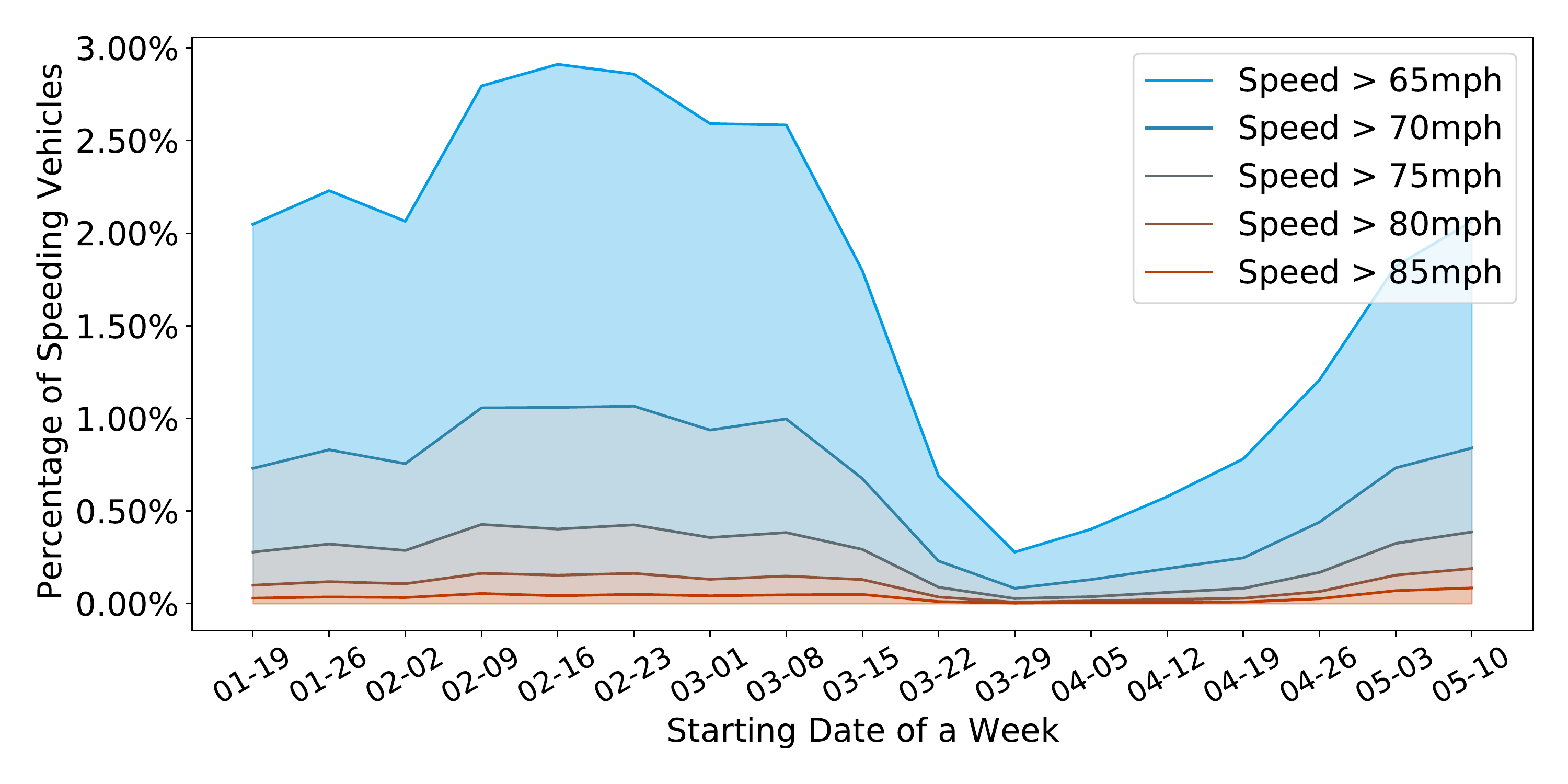}
         \caption{I-5, Northbound, Milepost from 154 to 181}
         \label{fig:residual_dow_pems}
     \end{subfigure}
     \hfill
     \begin{subfigure}[b]{0.48\textwidth}
         \centering
         \includegraphics[width=\textwidth]{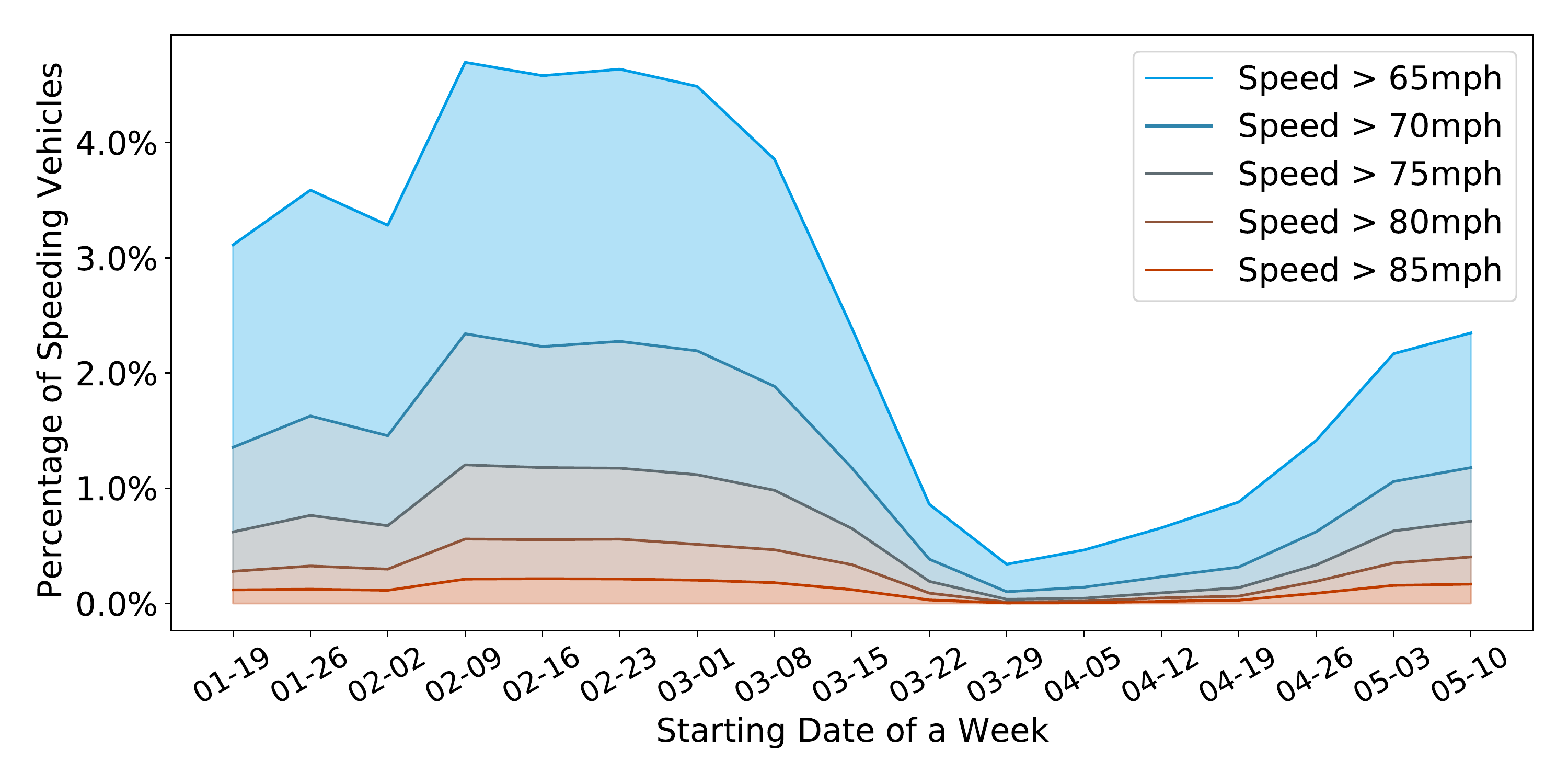}
         \caption{I-5, Southbound, Milepost from 154 to 181}
         \label{fig:residual_hod_pems}
     \end{subfigure}
     \\
     \begin{subfigure}[b]{0.48\textwidth}
         \centering
         \includegraphics[width=\textwidth]{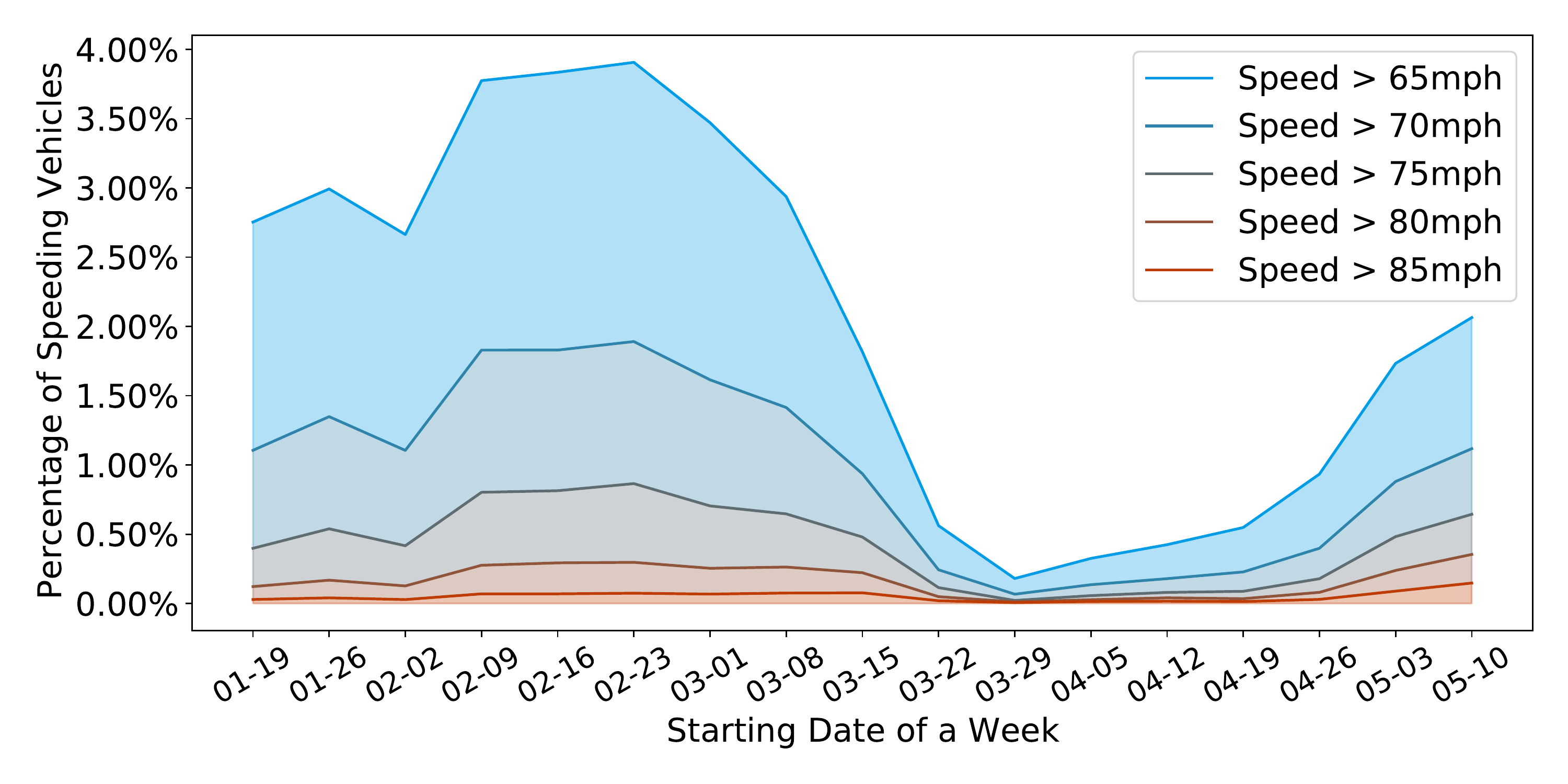}
         \caption{I-405, Northbound, Milepost from 0 to 30}
         \label{fig:residual_dow_metr}
     \end{subfigure}
     \hfill
     \begin{subfigure}[b]{0.48\textwidth}
         \centering
         \includegraphics[width=\textwidth]{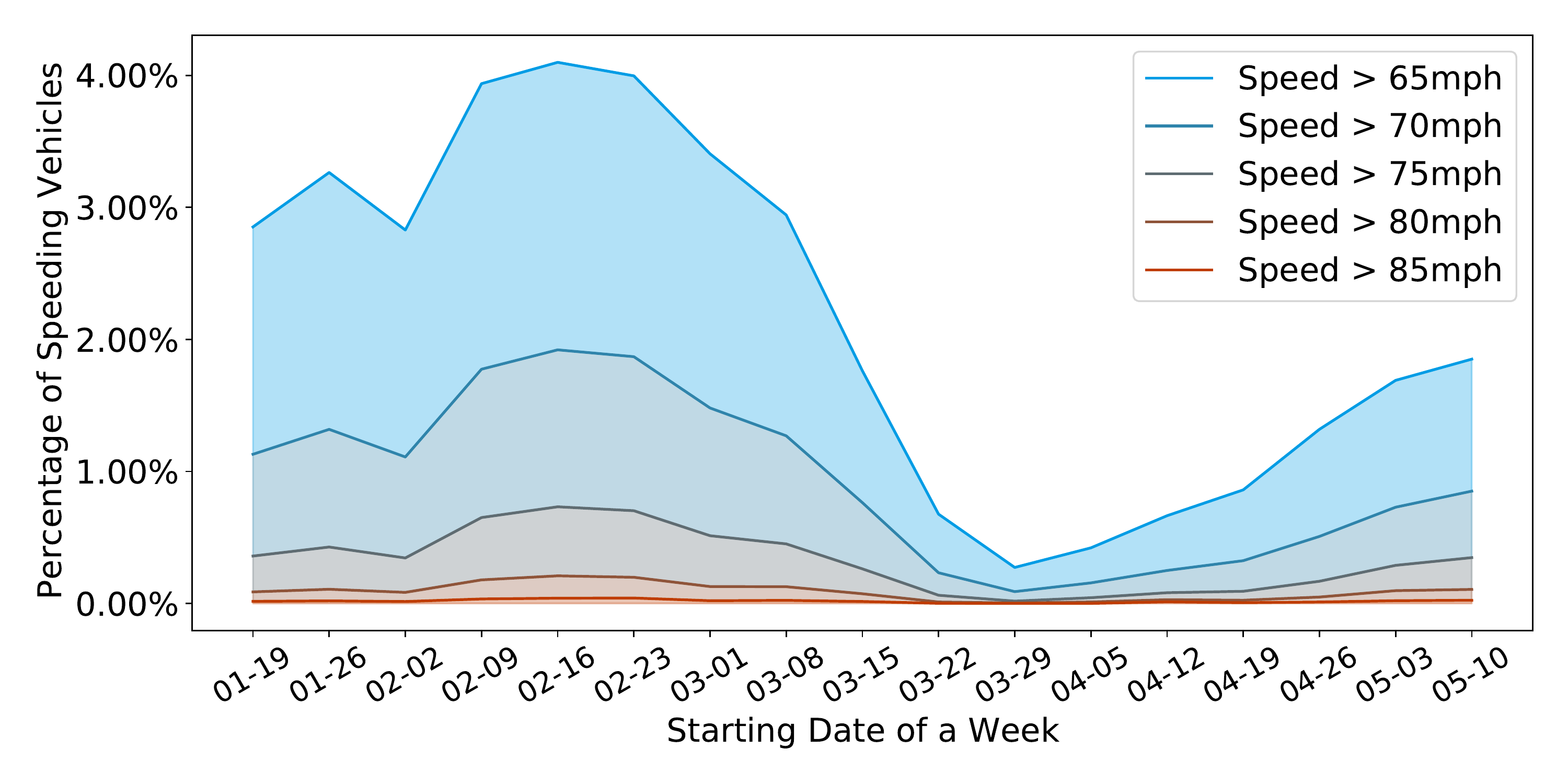}
         \caption{I-405, Southbound, Milepost from 0 to 30}
         \label{fig:residual_hod_metr}
     \end{subfigure}
    \\
     \begin{subfigure}[b]{0.48\textwidth}
         \centering
         \includegraphics[width=\textwidth]{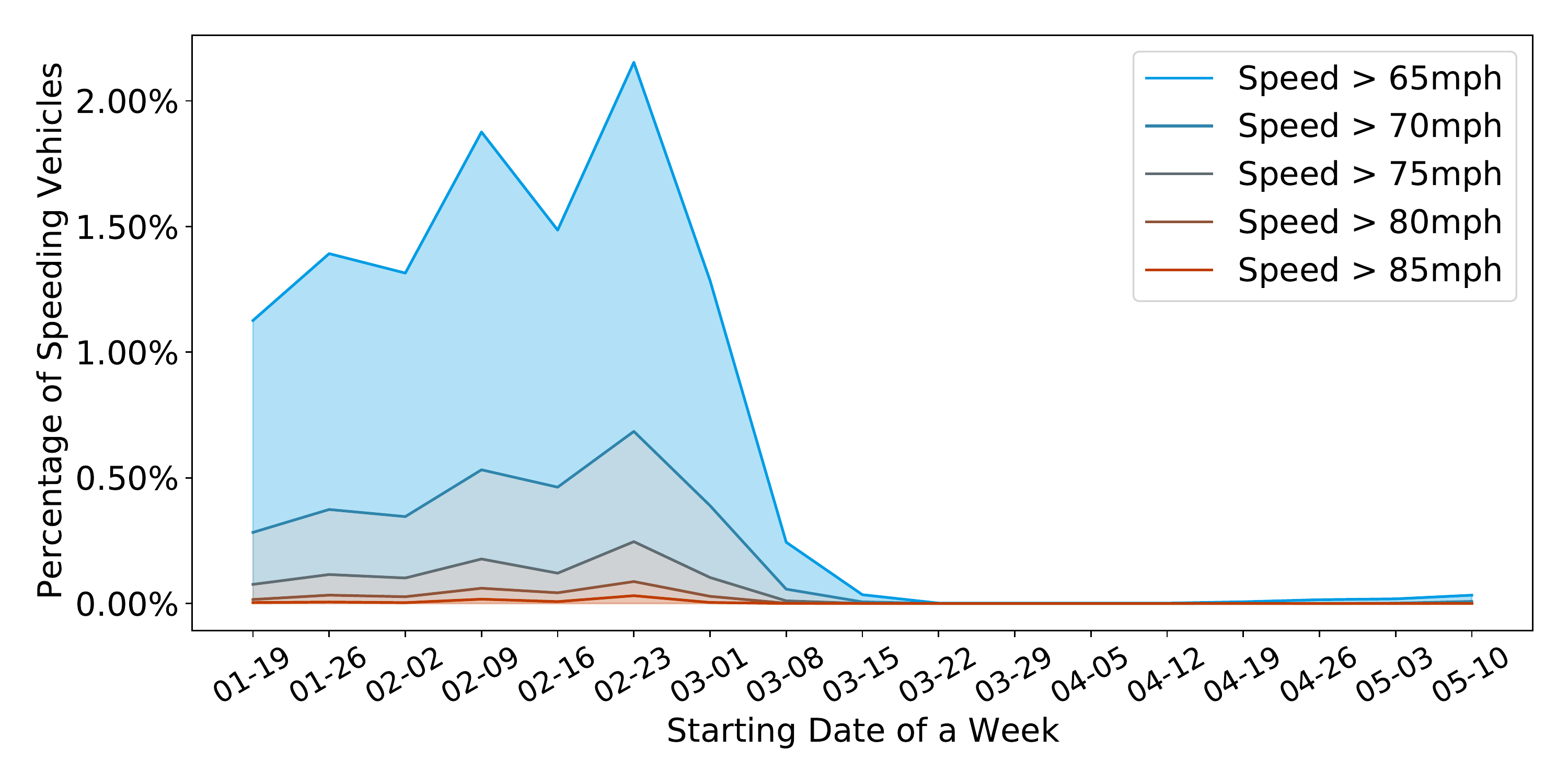}
         \caption{SR-520, Eastbound, Milepost from 0 to 12}
         \label{fig:residual_dow_inrix}
     \end{subfigure}
     \hfill
     \begin{subfigure}[b]{0.48\textwidth}
         \centering
         \includegraphics[width=\textwidth]{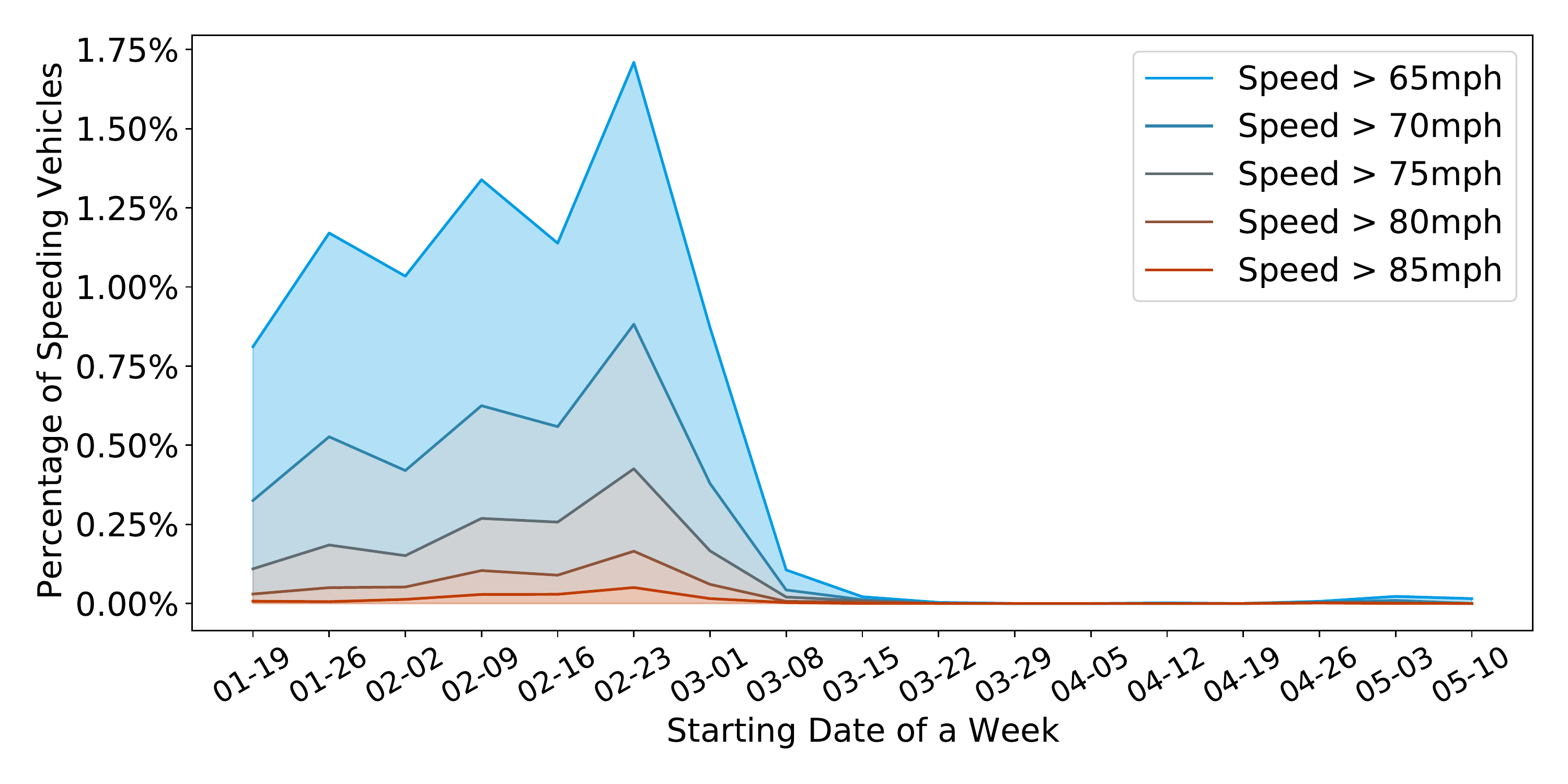}
         \caption{SR-520, Westbound, Milepost from 0 to 12}
         \label{fig:residual_hod_inrix}
     \end{subfigure}
     
    \caption{Percentages of speeding vehicles on major freeways in the Greater Seattle area before - and during - COVID-19.}
    \label{fig:speeding}
\end{figure*}

Because travel demand during COVID-19 has changed, the traffic scenarios for drivers has also changed. Thus, an individual's driving behavior might also be influenced by COVID-19. News \cite{SULLIVAN2020, Kaji2020} reported a rise in speeding in many States in the U.S. These conclusions in the news are traditionally drawn from officers' observations or based on the statistics of fatal crashes. In this section, we analyze and compare the speeding rates before, and during, COVID-19.

Here, the percentage of speeding vehicles on a specific corridor or the road network based on the lane-level loop detector data is calculated with one-minute time interval using the following equation
\begin{equation}
    Speeding Rate = \frac{\sum_{V_i > V_{TH}} Q_i}{\sum Q_i}
\end{equation}

\noindent where $V_i$ and $Q_i$ are the detected speed and volume of a loop detector and $V_{TH}$ is the speeding threshold. Figure \ref{fig:speeding} shows the percentages of speeding vehicles on three major freeways, I-5, I-405, and SR-520, in the Greater Seattle area during each week before, and during, COVID-19 (from Jan. 19 to May. 10). Each sub-figure plots the time-varying percentage curves with different speeding thresholds ranging from 65 miles per hour (mph) to 85 mph. The percentages of speeding vehicles on I-5 and I-405 on both directions, as shown by Figure \ref{fig:speeding} (a) to (d), have similar trends. 
The speeding rates reached a peak before COVID-19 around mid-February. In March, the speeding rates gradually reduced and reached their lowest points at the end of March. Since then, the speeding rates have started to increase. Ultimately, Figure \ref{fig:speeding} clearly shows that speeding rates during COVID-19 were significantly less than those before COVID-19. The reduction in speeding rates on SR-520, as shown in Figure \ref{fig:speeding} (e) and (f), are even dramatic than those for I-5 and I-405  after the stay-home order, as they suddenly reduced close to zero and did not rebound like the percentage curves on I-5 and I-405. 

It is important to note here, that when using loop detector data, the speed values of different vehicles within the one-minute time interval observed by a loop detector are averaged. Thus it is possible that one vehicle passing a detector may be driving at excessive speeds, but the observed averaged speed of the detector within one minute does not capture that outlier. While a small number of drivers may have driven at much higher speeds, the analysis results indicate that the overall speeding rate of freeways in the Greater Seattle area during COVID-19 is noticeably less than that before COVID-19.

\section{Conclusion and Discussion}

In this work, the traffic performance score to measure network-wide traffic performance is proposed. Based on real-time loop detector data and other data sources, a traffic performance score platform to display both network-level and segment level freeway traffic performance in the Greater Seattle area is developed. Based on real-time traffic data and the calculated traffic metrics, the impact of COVID-19 on traffic before and during COVID-19 are analyzed. The results show that both TPS and VMT have similar patterns. Namely, that the urban traffic has been greatly influenced by COVID-19, especially after the governors stay-home order was announced. The road segment-level travel demand changes and driver behavior changes are also measured, which reflects the ways COVID-19 is reshaping urban mobility. 

The traffic patterns, travel demands, and driving behaviors in different segments of the region have greatly changed following the outbreak of the pandemic. These impacts could be considered temporary. However, many large, regional employers, such as technology companies, have begun planning to allow employees to work from home permanently. If COVID-19 ultimately results in a work-from-home revolution as many are hypothesizing, urban mobility will likely not return to pre-COVID-19 patterns. Thus, the newly developed methods described in this work will be vital for quantifying changing traffic patterns into the foreseeable future. Because the TPS platform is built for scalability, it can easily be extended to cover more cities and regions across the country and the globe, and can facilitate more comprehensive analysis functions. It will be an important tool for agencies as they consider new policy directions moving forward.

\section{Acknowledgements}

This work is support by the Pacific Northwest Transportation Consortium (PacTrans) and the Smart Transportation Applications and Research Lab (STAR Lab) at the University of Washington. We would like to recognize the help from the Washington State Department of Transportation (WSDOT) for providing the loop detector data.

\bibliographystyle{unsrt}  
\bibliography{references}  






\end{document}